# Control and sorting of eccentric all-dielectric core-shell nanoparticles using two counter-propagating plane waves

**Ricardo Martín Abraham-Ekeroth[1-4], Marcelo Lester[1-3]**

[1] Instituto de Física Arroyo Seco (IFAS), Pinto 399, 7000 Tandil, Argentina
[2] Centro de Investigaciones en Física e Ingeniería del Centro de la Provincia de Buenos Aires (UNCPBA-CICPBA-CONICET), Campus Universitario UNCPBA, 7000 Tandil, Argentina
[3] Consejo Nacional de Investigaciones Científicas y Técnicas (CONICET), Godoy Cruz 2290 (C1425FQB) Buenos Aires, Argentina
[4] Universitat Jaume I, Av. Vicent Sos Baynat, s/n, 12071 Castelló de la Plana, Spain

E-mail: abraham@uji.es

## Abstract

Many typical nanoscale structures consist of dielectric nanoparticles with an inevitable oxide-generated coating around them. Depending on the fabrication techniques, these coatings may not be homogeneous, and their distortion can cause problems in the applications of such systems. Based on finite element simulations, inhomogeneous core-shell nanoparticle systems are numerically studied when illuminated with two counter-propagating plane waves in the optical range. It is found that the electromagnetic field distortions caused by the inhomogeneous system under Mie resonance conditions allow the system to exhibit controllable one-directional impulsion and rotation, which mainly depends on the offset between the core and shell. The overall geometry and composition of the system also dictate the type of resonance being excited. Overall, this "photonic thruster" effect consisting of an accelerating and spinning projectile would provide stability to particle movement and additionally establish a method to distinguish inhomogeneous from homogeneous particles. The method can be scaled to a wide range of nanoscale dielectric particles. Thus, the results are useful for detecting defects in nanosystems with a simple concept and may open avenues for improving nanoparticle synthesis methods.

## 1. Introduction

The ability to manipulate matter with light is nowadays an accessible tool since many applications have proven to be extremely useful, like microscopy [1–4], light sail development [5], living cells' characterization, manipulation, and surgery [6–11], microrheology [12,13], cancer theragnosis [9,14], cooling of gases by laser radiation [15,16], tracking of bacteriae and viruses [17], and remote sampling [18], along with others.

Furthermore, the core-shell nanoparticle systems have become essential in the nanoscale for several reasons. They can serve as multifunctional sensors or smart drug-delivery vehicles in biology and medicine [19]. They can be regarded as a simplified form of metamaterials, as they combine two or more materials with distinct properties within hybrid nanoparticles, thus offering the potential to enhance or customize their characteristics [20,21]. Protecting the core/s with shell layers that may even be incompatible with each other is one of the best advantages of this type of structures over the others [22]. Currently, precise synthesis methods can imbue these nanoparticles with highly advantageous properties such as thermal stability, solubility, and low toxicity [21–23]. Consequently, these attributes render the particles suitable for a wide range of applications, including drug delivery and detection [21], bioimaging, sensors [24,25], solar cell development, removal of heavy metals [22], and surface modification [23], to name just a few.

In particular, the use of all-dielectric core-shell nanoparticles can represent specific advantages over their counterparts due to simple reasons[24,25]. For example, all-dielectric nanoparticles can support electric and magnetic volumetric resonances (Mie resonances) [26,27], which involve hot spots within the materials at their lowest-energy modes [28]. This is distinct from nano systems that support surface resonances like surface plasmons, which create hot spots near tips or surfaces, potentially making the particles invasive in their surroundings [29,30]. Consequently, an appropriate design of all-dielectric coated particles can lead to lower energy loss than designs using metallic nanoparticles. For instance, the dielectric function of the shell layer can be optimized for the miniaturization of dielectric capacitors in electronic and electrical systems [31,32]. All-dielectric particles can also exhibit enhanced high-index sensitivity for sensing applications [33]. Even a variety of natural systems like a living cell or active matter can sometimes be conceived as all-dielectric core-shell systems [34,35].

However, although many core-shell nanoparticle types are on demand [36–39], the precise control of the particle fabrication is still far from being achieved. For example, the yolk–shell structure has recently shown excellent physical and chemical properties, thus showing great potential in many fields, such as lithium-ion batteries, drug delivery, nanoreactors, and catalysis [40,41]. This sort of structure can even represent particles with movable or isolated cores to gain some flexibility, including volume expansion. Nonetheless, this and many other kinds of complex core-shell systems may present deficient ripening processes or undesirable final structures, thus failing their application purposes [42]. Moreover, the healing of structures with defects has become an active research field to improve the efficiency of fabrication methods, especially those based on ripening processes or techniques involving dramatic volume change that eventually leads to particle fracture. [40,43]. On the other hand, breakable shells can become also essential [44]. Therefore, core-shell flexible designs with controllable inhomogeneities allow for multifield sensing or mass/energy storage for industrially scalable products [20,45].

In this regard, efficient separation or sorting methods are essential for the development of manipulation of nanoscale devices required for high demand manufacturing purposes.

Many classes of procedures have been established for particle sorting, a few involving complex physical-chemical processes [21,23,46]. Others are based on purely physical phenomena, such as optical forces. [12,47–49]. However, most of these methods become difficult to apply because they involve heavy-duty instruments, or lack control of monodispersity, stability, repetitiveness, or result economically inviable [20,36,37,39,50].

In this work, a method to sort core-shell nanoparticles is presented based on the resonant properties of all-dielectric systems and the symmetry breaking that occurs when there are defects in the particles, like inhomogeneous shells. As the illumination is composed of two counter-propagating optical waves with equal circular polarization, the scattered fields of the system turn out very sensitive to any rupture of the illumination symmetry. This numerical study focuses on the example with a silicon core and silica shell as the latter can exist as a byproduct of the oxidation of the core [51–53]. Silicon is usually regarded as one of the best cores in core-shell structures for energy storage [40]. In addition, its optical properties (high contrast) make it suitable for the excitation of electric and magnetic resonances [28]. In any case, the results in this paper remain general and independent of size, specific geometry, and constitution of the particles. A previous study on similar structures can be found in Ref. [12]. However, such work has just reported particle rotation effects on small systems. Here, a wide range of sizes is studied, and another important result is presented, namely, the net propulsion effect when inhomogeneous shells are present in the sample.

## 2. Methods

Figure 1 shows the geometry under study. Two counter-propagating plane waves of the same handedness, namely, left-handed circular polarization (LCP), constitute the incident field upon the system. The scattering body is composed of a silicon core (Si)-silica shell (SiO2) nanoparticle, whose core being in the displaced position $\vec{d}$, different in general from the whole particle centre in the origin of the cartesian coordinate system. $\vec{d}$ can also be characterized by the angles and radius defined in spherical coordinates, i.e., $(d, \varphi, \theta)$. The following dimensionless variable definitions were calculated from the electromagnetic fields which are solution of the problem,

$$Q_{abs} = \frac{4}{A_p c \varepsilon_0 |E_0|^2} \int_{V_p} Re\{\vec{J} \cdot \vec{E}^* - i\omega \vec{B} \cdot \vec{H}^*\} dV, \qquad (1)$$

$$Q_{sca} = \frac{4}{A_p c \varepsilon_0 |E_0|^2} \oint_{S_p} Re\{\vec{E}_{sca} \times \vec{H}_{sca}^*\}\breve{n} ds, \qquad (2)$$

Where $Q_{abs}$ and $Q_{sca}$ are the optical efficiencies for the particle in vacuum, $V_p$ [$S_p$] is the particle volume [surface], and $A_p$ is the area of the particle transversal to the illumination. $\breve{n}$ is the versor normal to the surface $S_p$.

The total time-averaged force $\langle \bar{F} \rangle$ and torque $\langle \bar{N} \rangle$ exerted on the particle can be calculated using the surface integrals,

$$\langle \bar{F} \rangle = \frac{1}{2} \oint_{S_p} Re\{\langle \bar{\bar{T}} \rangle . \check{n}\} ds, \qquad (3)$$

$$\langle \bar{N} \rangle = \frac{1}{2} \oint_{S_p} \check{n} \times Re\{\langle \bar{\bar{T}} \rangle . \check{n}\} ds, \qquad (4)$$

where $\langle \bar{\bar{T}} \rangle$ is the time-averaged Maxwell stress tensor, whose form for time-harmonic fields is given explicitly by

$$\langle \bar{\bar{T}} \rangle = \varepsilon_0 \bar{E} \otimes \bar{E} + \mu_0 \bar{H} \otimes \bar{H} - \frac{1}{2}(\varepsilon_0 |\bar{E}|^2 + \mu_0 |\bar{H}|^2) \bar{\bar{I}}, \qquad (5)$$

where $\otimes$ is the outer product and $\bar{\bar{I}}$ is the unit tensor. In this way, all the corresponding dimensionless variables are defined in a way similar to that for Ref. [54] for suitability; the efficiencies

$$Q_{abs} = \frac{4}{A_p c \varepsilon_0 |E_0|^2} \int_{V_p} Re\{\bar{J}. \bar{E}^* - i\omega \bar{B}. \bar{H}^*\} dV, \qquad (1)$$

$$Q_{sca} = \frac{4}{A_p c \varepsilon_0 |E_0|^2} \oint_{S_p} Re\{\bar{E}_{sca} \times \bar{H}_{sca}^*\} \check{n} ds, \qquad (2)$$

are equally defined, and the mechanical variables are defined as $F_{rad} = \frac{\langle F_z \rangle}{\omega_E^{tot} A_p}$, $\bar{N}'_{spin} = \frac{\langle \bar{N}_{spin} \rangle}{\omega_E^{tot} V_p}$. The electromagnetic fields $\bar{E}$, $\bar{B}$, $\bar{H}$, $\bar{E}_{sca}$, $\bar{H}_{sca}$ and their conjugates were thus calculated by simulations based on a finite element method (FEM) performed by the authors on a Python code and tested extensively (no details shown here for brevity).

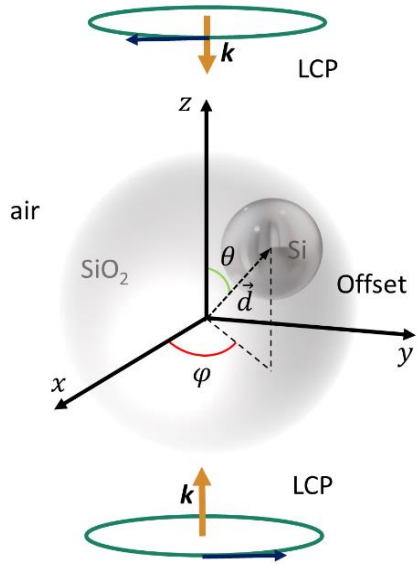

Figure 1: (Colour online) Definition of geometry, materials, and illumination configuration. The offset's distance and angles are defined by the vector $\vec{d}$, following the usual spherical coordinate system.

## 3. Results and Discussion

The discussion centres around a main illustrative example for a 300 nm particle with a 100 nm core and then briefly extends to other ranges in the nanoscale. The characteristic spectra of optical efficiencies are discussed in (3.1). Then, the study deals with core offsets parallel to the illumination axis, where a resonant "thruster" effect is obtained, see (3.2). A broader discussion about arbitrary offset directions for a particular offset value is made in (3.3). Finally, the study of (3.2) is extended in subsection (3.4) to include the radiation pressure at other nanoscale particle sizes.

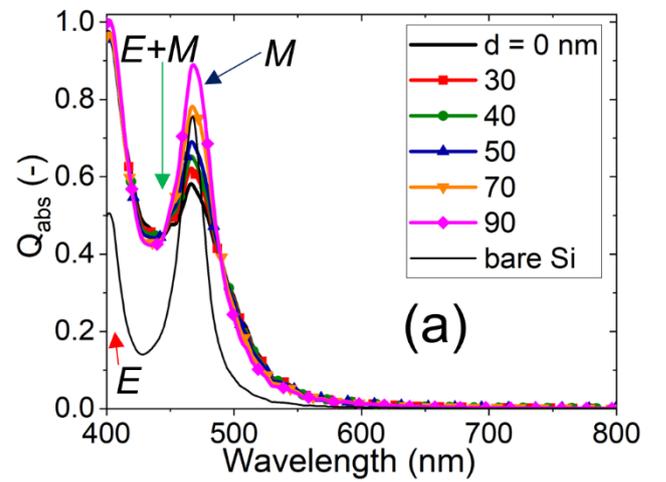

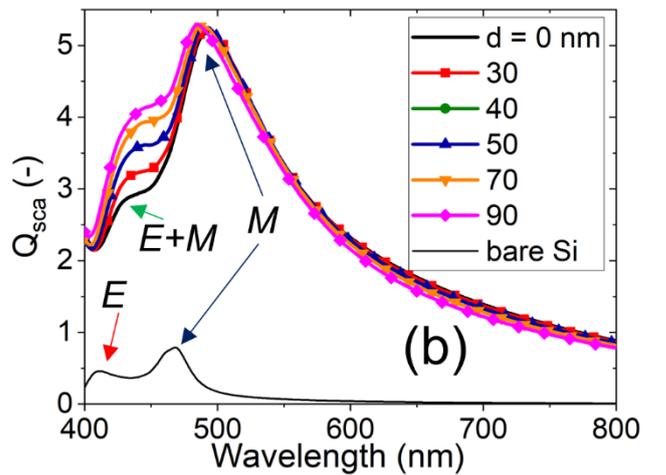

Figure 2: (Colour online) Spectra of optical efficiencies for several offset values. (a) [(b)] absorption [scattering]. Black solid [thin] line for the homogeneous shell (zero offset) [bare silicon, no shell]. E and M labels indicate the electric and magnetic dipolar resonances derived from near fields around the silicon core. The label E+M

means a hybrid resonance due to the overlapping between E and M modes.

### 3.1. Optical behaviour: efficiencies.

Figure 2 a [b] shows the spectra of absorption [scattering] efficiency for several values of the offset magnitude along the z direction. Black thick [thin] corresponds to a concentric core-shell system [bare core]. Note that the particle varies their optical intensities, but almost no resonance shifts occur. Moreover, the nature of the Mie resonances can be deduced from an analysis of the curves by comparison with the results for a bare core and the behavior of the near fields (not shown for brevity)[26]. As a result, just two volume modes appear for the illustrative case with a relatively small size 300-100 nm; the lowest energy (longest wavelength) mode is the magnetic dipole (M), while the highest energy (shortest wavelength) is the electric mode (E). Despite this classification, when the core is somehow surrounded by the $SiO_2$ shell, the overlapping between the M and E modes is so strong it makes the resonance a hibrid mode (labeled E+M in Figure 2), as a product of the interaction between the core and the shell.
Although the efficiencies respond to an alteration in the system, such as the core-shell offset, they are not enough to completely characterize the system nor identify in which direction the offset is located, especially under symmetric illumination conditions (see Figure 1). On the other hand, the mechanical inductions on the particle are expected to contain more information of the near-fields [28,55,56].

### 3.2. Resonant "thruster" effect for parallel offsets.

As the two optical modes labeled E+M and M proved to be more sensitive to the offset's variation, Figure 3 shows the forces and torques exerted on inhomogenous particles as a function of the z-offset for three resonance locations, namely, 440 nm (E+M mode), 490 (M), and 800 nm, this latter constituting an out-of-resonance example. Surprisingly, a net radiation force (Figure 3a) and torque (b) appear due to the multiple interactions between the core and shell as soon as the offset ceases to be zero. As can be seen from the curves, the phenomenon is resonant: it is almost zero out of resonance and increases as the intensity of the resonance does. Moreover, it results proportional to the offset distance from the particle's centre. In particular, there is a crossing point between the curves for 440 and 490 nm in the torque so that the maximum is achieved for the highest possible value of offset under M resonance.

Note that both mechanical reactions occur simultaneously and in a preferential direction since the offset is assumed in the positive z-axis. This "thruster" phenomenon provides the particle with both propulsion and spin, therefore self-stabilizing its movement as in a rocket system. Remarkably, this effect represents a potential detection mechanism for identifying inconsistencies in particle fabrication. The particle itself will be compelled to move and spin in the same direction as anomalous particles, akin to how they behave in a velocity-mass selector.

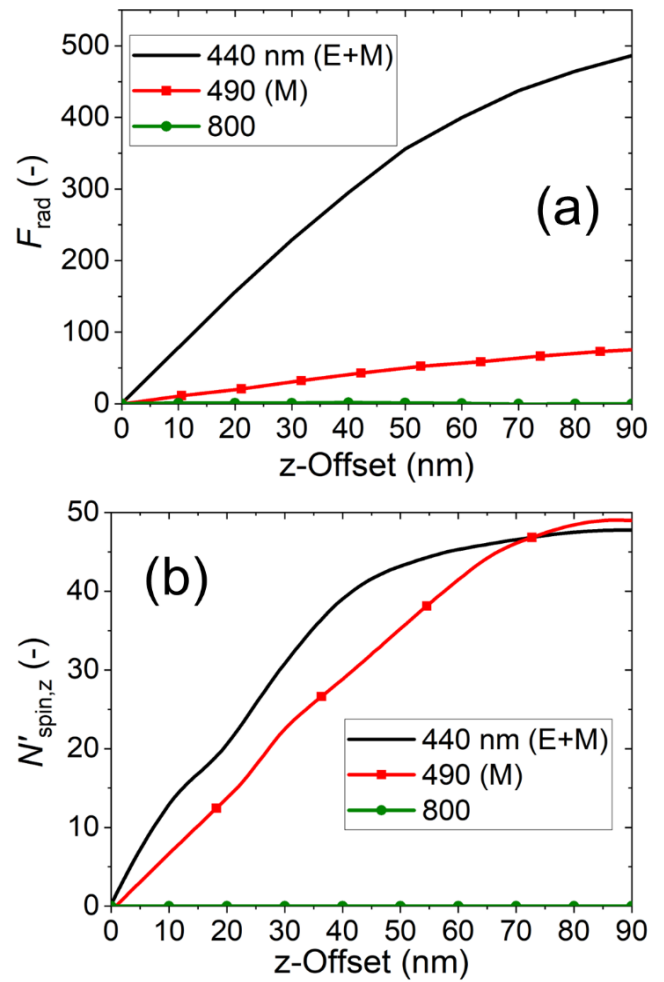

.
Figure 3: (Colour online) Force (a) and Torque (b) as a function of the offset ($d$) along z for three different wavelengths: Electric (E) and magnetic (M) resonances, and the non-resonant wavelength at 800 nm. The wavelength locations of the resonances correspond approximately to those for $Q_{sca}$ in Figure 2.

Following the study, a natural question arises: How will the particle move if it is arbitrarily oriented with its anomaly in a direction different from the illumination's? This is the subject of the next subsection.

### 3.3. Mechanical predictions for arbitrary offsets.

Figure 4 and Figure 5 show the net force [torque] (a) [(b)] exerted on the particle's *centre of mass* (CM) for the resonances E+M, at 440 nm, and M at 490 nm, respectively.

In other words, each arrow begins at the exact location of the CM of the particle and is appropriately scaled. The arrows in Figure 4a [b] are scaled by a factor $2.5 \times 10^{-3}$ [$7.5 \times 10^{-3}$]. In Figure 5, the arrows in both panels are scaled by a factor of 0.01. For a better visualization, the red [blue] arrows distinguish the movement along the positive [negative] orientation in the z-axis. The three-dimensional perspectives of these maps are also chosen carefully to allow the reader to better comprehend how the particles would move.

In general, all the maps show mirror symmetry around a plane of high symmetry, i.e., $z = 0$. As can be seen in the arrow maps, Figures 4 and 5, the tendency of the particle to go up or down in the illumination direction (along the z-axis) is very evident. Those positions for the particle's CM out of the plane $z = 0$ imply a net radiation force in the direction of the anomaly, Figure 4a. In a cluster of *dispersed* particles, this phenomenon would allow, in principle, a clear separation of homogeneous particles, namely, particles with perfect shells, from anomalous ones.

The effect becomes more complex as some rotations are induced simultaneously (panels b in Figures Figure *4* and Figure *5*). The mirrored, vortex-like shape of the torque maps in these panels reveals that the particle's CM would try to self-locate on the z-axis of symmetry if it is out of the plane $z = 0$.. The reader is reminded that the torque arrows show the tendency of the CM to rotate or spin *around* the direction given by these vectors or, equally, in local planes perpendicular to them. Notice that the force arrows are larger around the "equator regions", which means a higher relative magnitude than around the poles. On the contrary, the torques are greater in the regions between the equator and the poles. This fact supports the hypothesis that the particle finds more stability when its CM is along the z-axis.

By comparing Figures Figure *4* and Figure *5*, it is evident that the overlapping of modes is also present in the near fields that produce the mechanical results. However, the relative magnitude of the force arrow around the polar regions is higher for the E+M than for the M mode, which gives the map in Figure 4a more aspect of a "balloon" than in Figure 5a.

There is another phenomenon that is worth mentioning in Figure 4 and Figure 5. In the equator CM positions, the resultant force would push the particle in a direction opposite to its anomaly. This pulling-force "circle" without rotation around the equators highlights the symmetry conditions of the illumination and produce instability for the particle when its CM lies on the $z = 0$ plane.

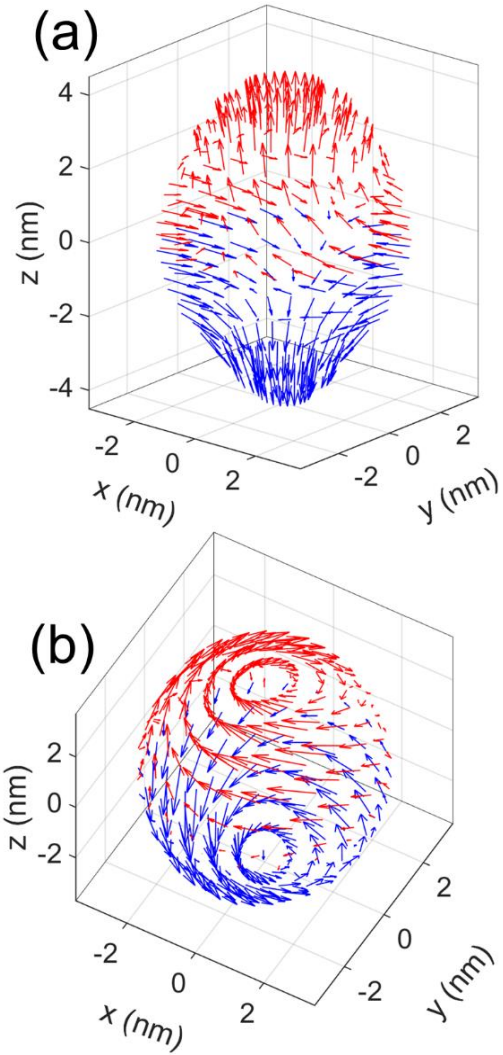

Figure 4: (Colour online) Net mechanical variables for the particle's centre of mass (CM) for several values of offset angles $\varphi$, $\theta$ at E+M resonance, i.e., 440 nm. Each arrow is the net (a) [(b)] force [torque] at the position $(x, y, z)$ of the CM. Red [Blue] arrows for $z>0$ [$<0$] position of the CM.

### 3.4. Variation of the radiation pressure with the size. Study of the size dependence.

Finally, with the aim to complete a parametric study, let's vary the size of the system beyond the example 100-300 nm when the offset is kept again along the z-axis. For conciseness, the results correspond to a set of particles with geometric similarity, as follows. If the minimum distance $\delta$ between core and shell is defined as $\delta = R_b - R_a - d$, where $R_a, R_b$ are respectively the core and shell radii, the scales $\delta/2R_a$ and $\delta/2R_b$ are kept constant while varying the particle's size.

## (a)

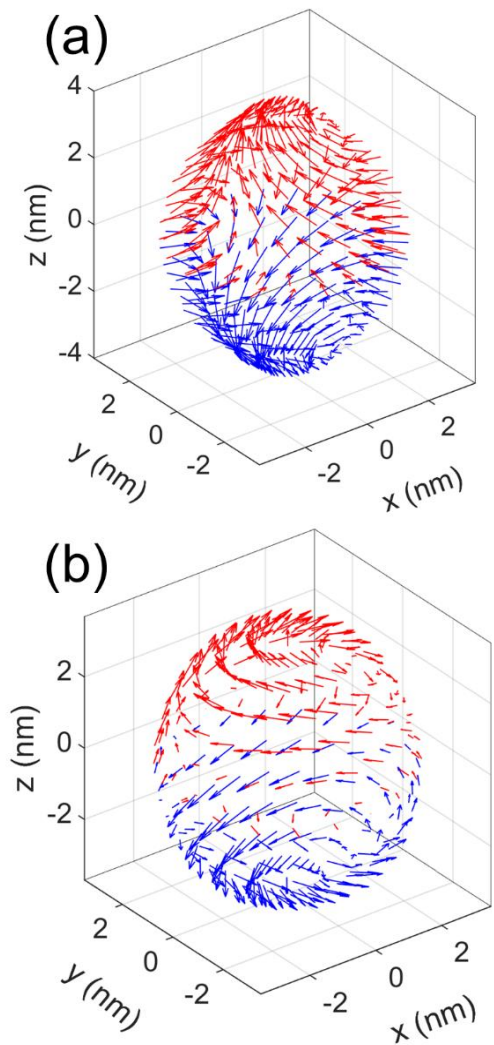

## (b)

Figure 5: (Colour online) Idem Fig. 4 for the magnetic (M) resonance, at 490 nm.

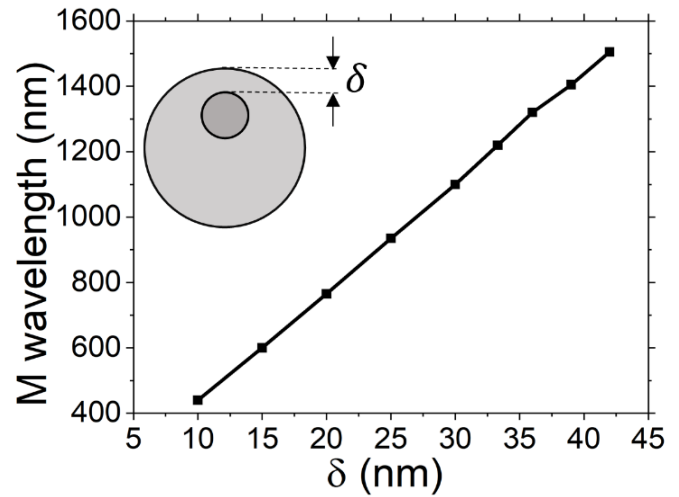

Figure 6: (Colour online) Study of the radiation force exerted on the particle as a function of the size for particles with geometrical similarity. The black line is the location of the lowest-energy magnetic resonance as a function of the minimum distance between the core and the shell, $\delta$.

## 4.  Conclusion

A method of detection of particle anomalies/particle sorting is proposed from a simple optical illumination by taking profit of the resonant properties of all-dielectric core-shell nanoparticles. In a broad range of nanoparticle sizes, the lowest-energy volumetric Mie resonances induce strong anisotropic mechanical inductions on the whole particle, such as unbalanced radiation pressure and particle rotations. These results are based on the symmetry breaking produced by inhomogeneous shells under symmetric illumination. The symmetry breaking induces anisotropic forces from the electromagnetic interaction between the core and the shell of the systems. In particular, the mechanical variables carry the information of that interaction through their dependence on the near fields. Remarkably, the relatively strong dipolar magnetic resonance (lowest-energy mode) sets an acceptable linear dependence of the optomechanics on the size of the particles, which allows for complete particle discrimination from inhomogeneous to homogenous shells. As the forces induced also grow with increasing offset values for a fixed size, the concept is capable in principle to distinguish degrees of anomalies in a range of values for the eccentricity. The fact that torques also arise from the induction would also matter in other applications different from particle selectors such, for instance, nanoprinting, nanoscale measurements of viscosity, and light sail propulsion, among others.

The line shown in Figure 6 corresponds to the radiation pressure exerted on the particle as a function of $\delta$. The analysis of the torques is avoided for brevity. The line points were obtained by getting the maximum of the spectra of forces at the location of the M dipolar resonance, or lowest-energy mode, since several multipolar resonances are excited when the particle size increases. Noticeably, a linear trend is obtained as $\delta$ varies, which would help identify the particles or calibrate the device if the method of particle separation were applied.

Note that no resonance appears below the value $\delta = 10$ nm, provided that such a small core cannot support Mie resonances of volume type as to continue the trend of geometric-similar systems. In any case, smaller values of $\delta$ would result in unrealistic since non-local corrections to the constitutive parameters are likely to appear due to quantum effects.

### Acknowledgements

The authors would like to thank CONICET and IFAS-UNCPBA for their funding and workplaces to allow for discussions on the topic. R.M.A.-E. also acknowledges partial support by the DYNAMO project (101046489), funded by the European Union, but the views and opinions expressed are, however, those of the authors only and do not necessarily reflect those of the European Union or the European Innovation Council. Neither the European Union nor the granting authority can be held responsible for them.

**Conflicts of interest**

The authors declare that the research was conducted in the absence of any commercial or financial relationships that could be construed as a potential conflict of interest.